\documentclass[10pt,twoside]{article}
\usepackage[T1]{fontenc}
\usepackage[applemac]{inputenc}
\usepackage{amssymb,amsmath}
\usepackage[]{graphicx}
\def\volnum{Manuscrit}

\usepackage{aflbcours}
\pdebut{1}

\title{Did Louis de Broglie miss the discovery of the Schr\"odinger equation? }
\titleshort{Did Louis de Broglie miss the discovery of the Schr\"odinger equation? }
\author{Aur\'elien Drezet$^{(1)}$}
\authorshort{A. Drezet}
\address{(1) Univ. Grenoble Alpes, CNRS, Institut N\'{e}el\\
 F-38000 Grenoble, France}
\begin{document}
\maketitle

\vskip 1cm
\begin{quote}
\textit{La nouvelle dynamique du point mat\'eriel libre est \`a l'ancienne dynamique (y compris celle d'Einstein) ce que l'optique g\'eom\'etrique est \`a l'optique ondulatoire.}\\
Louis de Broglie, Quanta de lumi\`{e}re, diffraction et interf\'erences.  C.~R. Acad.~Sci. (Paris), \textbf{177}, 548--560 (1923) 
\end{quote}
\begin{abstract}
 In this note, we discuss a historical point regarding Schr\"odinger's discovery of the famous quantum wave equation in 1926 following de Broglie's fundamental works published in 1923-1925 regarding the introduction of matter waves. Drawing on the work of historians and personal analysis, we show that de Broglie was very close to the discovery of the Schr\"odinger equation (at least for the stationary one-electron problem).  
\end{abstract}


The title of this  note is deliberately provocative. In this anniversary year, when we celebrate Louis de Broglie's fundamental discovery of matter waves in 1923 \cite{debroglie1923a,debroglie1923b,debroglie1923c,debroglie1925th}, the question arises as to the developments that immediately followed this pioneering work. One of the questions that physicists and historians inevitably ask concerns Schr\"odinger's wave equation. 
How is it possible that Louis de Broglie missed the wave equation discovered by Erwin Schr\"odinger at the end of 1925 and published in the spring of 1926? \cite{Schrodinger1926} This question is indirectly addressed by historians V.V. Raman and Paul Forman, who in 1969 published an article entitled "Why Was It Schr\"odinger Who Developed de Broglie's Ideas? \cite{Forman} The authors answer the question of why Schr\"odinger's scientific and cultural background made him the most likely person to take the mathematical leap of putting Louis de Broglie's work into a (wave) equation. 
More recently, the great French historian Olivier Darrigol posed the question that interests us in a 2013 work entitled "A few reasons why Louis de Broglie discovered matter waves and yet did not discover Schr\"odinger's equation" \cite{Darrigol2013}. 

Darrigol's nuanced answer is worth reproducing here. He begins by recalling how de Broglie obtained his theory of matter waves and wave-particle duality (or rather harmonious coexistence) in 1923 by following an atypical and iconoclastic path. Indeed, based on his reading of the proceedings of the first Solvay Congress in 1911, de Broglie was quickly struck by the deep analogies between Fermat's principle for waves in the so-called "geometric" regime, on the one hand, and Maupertuis' principle in classical material point mechanics, on the other (at a mathematical if not physical level, this link had been known since the work of Hamilton and Jacobi in the 19th century). In connection with the "phenomenological" Bohr-Sommerfeld quantization rule imposed on the stationary orbits of atoms, de Broglie intuited a profound link between these different problems. This is not the place to recall the details of de Broglie's reasoning in 1923, which enabled him to define and characterize his matter wave accompanying the motion of any particle, as these will be recalled in greater detail in other contributions to this special issue of the annals celebrating "the centenary of the discovery of matter waves".

In his very rich article, Darrigol mentions an immediate link between de Broglie's work and Schr\"odinger's wave equation. He shows quite simply (I'll come back to this point later) that de Broglie could have defined a refractive index for his matter wave (or phase wave) as early as 1924, based on his thesis work \cite{debroglie1925th}. This local index $n_E(\mathbf{x})$, defined at the space point of coordinate $\mathbf{x}$ and written here in the relativistic regime corresponding to de Broglie historical approach, reads
\begin{equation}
n_E(\mathbf{x})=\sqrt{[(1-\frac{eV(\mathbf{x})}{E})^2-\frac{m^2c^4}{E^2}]} \label{eq1}
 \end{equation}
where $E$ is the energy for a particle with electric charge $e$ and mass $m$ moving in an electrostatic potential  $V(\mathbf{x})$.
This index is quite natural when one starts from de Broglie's phase-wave theory, which imposes the wave velocity as being
\begin{equation}
V_{wave}=\frac{E}{P}=\frac{cE}{\sqrt{[(E-eV)^2-m^2c^4]}}=\frac{c}{n_E}.
\end{equation}
which depends on the relativistic momentum $P/c=\sqrt{[(1E-eV)^2-m^2c^4]}$.
Of course, it can also be deduced from Hamilton-Jacobi theory.
By introducing the Helmholtz wave equation :
\begin{eqnarray}
\boldsymbol{\nabla}^2\Psi(\mathbf{x})+\frac{\omega^2}{c^2}n_E^2(\mathbf{x})\Psi(\mathbf{x})=0
\end{eqnarray} 
for a  stationary wave $\Psi(\mathbf{x})e^{-i\omega t}$ with  pulsation $\omega$, and using  the Einstein-de Broglie relation $E=\hbar\omega$ and the index formula Eq.~\ref{eq1}, we naturally find the relativistic Klein-Gordon stationary wave equation  
\begin{eqnarray}
\boldsymbol{\nabla}^2\Psi(\mathbf{x})+\frac{1}{\hbar^2c^2}[(E-eV(\mathbf{x}))^2-m^2c^4]\Psi(\mathbf{x})=0.
\end{eqnarray} 
Moreover, in the non-relativistic regime (writing $E=E_{NR}+mc^2$ with $|E_{NR}-eV|\ll mc^2$) we get:
\begin{eqnarray}
\boldsymbol{\nabla}^2\Psi(\mathbf{x})+\frac{2m}{\hbar^2}[E_{NR}-eV(\mathbf{x})]\Psi(\mathbf{x})=0
\end{eqnarray}
which is the well known Schr\"odinger stationary wave equation for a single particle in an external potential.

After these technical developments, Darrigol naturally asks why de Broglie's 1924 article lacks such reasoning.
The passage from Darrigol's article that I would like to reproduce here is as follows:

\begin{quote}
Why did not de Broglie follow this simple track? A first element of the answer is that, 
notwithstanding with his grand analogy between dynamics and optics, he was shy in adventuring beyond the approximation of geometrical optics. He focused on retrieving results of the received quantum theory such as the Bohr--Sommerfeld conditions, and he underplayed the more disturbing consequences of his concept of matter waves. Another possible obstacle to his developing a wave theory of matter was his conviction that both light and matter had a dual nature, implying the synchronous motion of waves and particles. This duality focused him on the interplay between waves and particles rather than on the search for a new wave equation. Thirdly and most importantly, de Broglie believed that the analogy between light and matter implied the electromagnetic nature of his matter waves. Consequently, he also believed that matter waves obeyed the d'Alembertian equation of electromagnetism. Direct evidence of this conviction is found in a note of 1925 \cite{debroglie1925} in which he describes the intrinsic oscillation of an electron in its rest frame as the stationary superposition of the retarded and advanced solutions of the d'Alembertian equation. The same heuristic principle, the analogy between matter and light, led de Broglie to the matter waves and prevented him from seeking a specific equation for these waves!\cite{Darrigol2013}\end{quote} 

Darrigol's deduction is particularly clear, but I'd like to qualify it here for several reasons. Firstly, de Broglie's 1925 note \cite{debroglie1925} is indeed fundamental, as it lays the foundations for what would later become de Broglie's double solution theory, a grandiose tantalizing unification of waves and particles in which the particle appears (no doubt inspired by Einstein's ideas on the photon) as a kind of mathematical singularity in motion of d'Alembert's wave equation. The theory of the double solution was then developed by de Broglie in more detail in 1926 and 1927 \cite{debroglie1926,debroglie1927}, based on the Klein-Gordon wave equation, and after a quarter-century of neglect was revived by de Broglie and his students in the 1950s-60s \cite{debroglie1956,Fargue}, based on the theory of nonlinear wave equations giving rise to so-called "solitons" or localized solutions.\footnote{For a modern discussion of this theory, see references \cite{Drezet2021,Drezet2023,Drezet2023b,Drezet2024}, which show that the double solution theory may be the future of physics (according to the present author). We note in passing that the double solution theory gave birth the same year to the "pilot wave" theory that de Broglie presented at the Solvay congress of 1927 \cite{debroglie1927,Valentini,debroglie1930}. This pilot-wave theory has survived to the present day, where it is known as the de Broglie Bohm theory or Bohmian mechanics \cite{Bohm1952} and constitutes a very serious interpretation of quantum mechanics.} What is remarkable in the 1925 paper, however, but not noted by Darrigol, is that it allows the existence of a phase wave to be derived in the absence of a $V$ potential:\footnote{In \cite{debroglie1925} de Broglie wrote (up to the sign difference) $e^{-i(\omega (t- z/V_{wave}}$ for a motion of the wave along the $z$ direction.}
\begin{eqnarray}
\Psi(\mathbf{x},t)=e^{-i(E t- \mathbf{P}\cdot\mathbf{x})/\hbar}
\end{eqnarray} which is a solution of the Klein-Gordon equation while the full wave, written  $u(\mathbf{x},t)$ by de Broglie, and which for him is the  only physical and real wave, obeys to the d'Alembert wave equation
$[\boldsymbol{\nabla}^2-\frac{1}{c^2}\partial_t^2]u(\mathbf{x},t)=0$.
However, although the 1925 article contains this phase wave and de Broglie's derivation, it does not mention that it is indeed a solution of the Klein-Gordon equation, which indeed had not yet been discovered! Schr\"odinger was the first to derive this Klein-Gordon equation in December 1925 and it was first published by Louis de Broglie, who rediscovered it in 1926 \cite{debroglie1926,debroglie1926b} based on a detailed analysis of the Hamilton-Jacobi equation.
So de Broglie was really very close to the wave equation as early as 1925, and this allows us to qualify Darrigol's assertion a little.

There are two other reasons why I believe that de Broglie was very close to discovering the wave equation, and that his failure to do so before Schr\"odinger was not just due to a lack of willpower. The first point is a personal anecdote. A few years ago, at a physics symposium, a retired French theorist who had attended Louis de Broglie's lectures in the 1950s in his youth confided in me that de Broglie had told him that Schrodinger had somehow "robbed" him of the discovery of the wave equation. Of course, this was just a remark made during a coffee-break lecture, but it disturbed me greatly, and if given any semblance of plausibility, it cast doubt on the idea of de Broglie having missed the wave equation through lack of conviction or desire.

Fortunately, there's another, slightly more objective reason to make sense of what I'm talking about, and that's Louis de Broglie's book Ondes et mouvements (Waves and motions) \cite{debroglie1925b}, published in 1926 but written in 1925. This text, which is in fact de Broglie's first book, is completely in line with the note \cite{debroglie1925} published the same year. However, the book goes further in its mathematical and physical developments. The book has not been fully appreciated by historians of science, and the fact is that de Broglie himself did everything in his power to ensure that it was forgotten. So, although it was in this magnificent book that de Broglie developed his first attempt at a double solution, it was never cited by de Broglie and his followers in the 1950s-70s as a seminal work. De Broglie will always instead refer to his 1927 article "Wave mechanics and the atomic structure of matter and matter" \cite{debroglie1927} as the true birth-act of his conceptions of the double solution and the pilot wave. The book has thus gone down in history and remained virtually ignored by the community.\footnote{See, however, recent developments by the author \cite{Drezet2023b,Drezet2024}, and Daniel Shanahan's article \cite{Shanahan2024} in this same issue.}

There are undoubtedly several reasons for this: Firstly, as mentioned by Darrigol, de Broglie's theory uses delayed and advanced waves, which implies a symmetrical causality running both forward from the past to the future and from the future to the past! It is therefore possible that de Broglie, sensing the highly speculative nature of his ideas, took fright and preferred to abandon them.\footnote{the present author believes, however, that this was a misjudgment on de Broglie's part concerning his own work \cite{Drezet2023b,Drezet2024}} Another hypothesis is perhaps that this work reminded him that he had indeed just missed Schr\"odinger's wave equation.

It's all speculative, of course, but de Broglie's book contains also a remarkable chapter on diffraction and scattering by particles of matter. This chapter of the book is important if only because it is one of the few places where de Broglie talks explicitly about matter-wave scattering in his early work. Indeed, everyone is familiar with the famous 1923 note \cite{debroglie1923b}, which contains the following short passage on matter particle diffraction:
\begin{quote}
This is where we may have to look for experimental confirmation of our ideas. \cite{debroglie1923b}
\end{quote}

It is also well known that it was this prediction that led to Davidson and Germer's great experimental discovery published in 1927, and its correct interpretation as confirmation of the existence of de Broglie's matter waves.
Moreover, in chapter 10 of his book Ondes et mouvements, entitled "Diffusion and Dispersion", de Broglie talks about Rutherford's famous experiment involving the Coulombian diffusion of charged particles by an atomic center. It is well known that Rutherford's experiment is regarded as a striking confirmation of the discrete, localized nature of the atomic nucleus. De Broglie proposes to treat the same scattering phenomenon qualitatively, using his phase-wave theory. The passage is so important that I reproduce it here in full:

\begin{quote}
From our point of view, we can consider the same problem from another angle. We have seen that the energy trajectories $E$ of a charge of value $e$ and proper mass $m$ coincide with the rays of a homogeneous wave of the classical type propagating in a medium of index defined at each point by the law
\begin{equation}
n_E^2=[(1-\frac{eV}{ h\nu})^2-\frac{m^2c^4}{h^2\nu^2}] \nonumber
 \end{equation}
$V$ is the electrostatic potential, $\nu$ the frequency $E/h$. If the potential is created by a charge $C$, it will be proportional at each point to the inverse of the distance to $C$. To obtain the scattering of a beam of electric particles under the influence of the charge $C$, it will suffice to study the deformation of an initially plane homogeneous wave as it propagates in a medium whose index varies according to the law indicated. The radii of this wave will be the trajectories of the deviated particles. If, initially, the moving particles are "associated with the same wave" in the sense we have given to this expression, the refracted wave will provide us with the phase distribution of the associated waves along the various rays.
Since particle deflection occurs in a small space around $C$, the deflected trajectories will appear at great distances to be straight lines passing through $C$; the scattering charge will therefore appear to be the center of rectilinear rays and the classical wave, which in the distance represents the particle distribution, will be a spherical wave very approximately. The distribution of amplitudes on the spherical wave will depend on the statistical law studied by Rutherford, since the square of this amplitude represents the average density of deviated particles. In short, if we limit ourselves to considering the homogeneous wave that statistically represents motion, and if we focus solely on describing phenomena far from the scattering center, we can say: under the action of the incident wave, the center emits a secondary spherical wave on which, moreover, the amplitude is not uniformly distributed. This statement reveals a deep kinship between the scattering of $\alpha$ or $\beta$ rays, for example, and the scattering of light conceived according to classical ideas. \cite{debroglie1925b}
\end{quote}

The text is very rich for the historian and demonstrates that de Broglie was ahead of his time on many concepts concerning quantum and wave theory. First of all, this text clearly mentions the medium index $n_E$ of our Eq.~\ref{eq1} (I've adapted de Broglie's notations but kept the frequency $\nu=\frac{\omega}{2\pi}$).  Also, de Broglie talks mainly about secondary and primary waves, and all this is just another form for the Born scattering formula
\begin{eqnarray}
\Psi=e^{i(Pz-Et)/\hbar}+ f(\theta)\frac{ e^{i(Pr-Et)/\hbar}}{r}
\end{eqnarray}   where $f(\theta)$ is the angular scattering coefficient  and  $r$ is the distance from the nucleus taken as origin.
The text also anticipates Born's statistical interpretation of $|\Psi|^2$, which should strictly speaking be called the de Broglie-Born statistical formula. De Broglie was therefore fully aware that his phase wave, which we note here as $ \Psi$, and which is only a part of the total wave $u$, represents particle motion (anticipating the pilot-wave of 1927). He was also the first to clearly perceive the statistical interpretation of such a phase wave, which here also becomes an amplitude wave.
But that's not all. Schr\"odinger's theory clearly shows that in general the angular distribution $|f(\theta)|^2$ can only recover Rutherford's formula in the semi-classical limit.\footnote{It is remarkable, however, that in the case of scattering by a Coulombian center of charge $C=-Ze$ ($e=-|e|$ being the electron's negative charge), the formula obtained for the classical Rutherford cross section coincides with the quantum scattering formula obtained by Mott in the non-relativistic regime. Here we find a specificity of the Coulombian potential that implies, among other things, that the Balmer formula obtained with Schr\"odinger's theory is identical to the semi-classical Bohr-Sommerfeld formula recovered by de Broglie.}
De Broglie doesn't mention this point here, but it raises some interesting questions. Could it be that de Broglie restrained himself in his speculations on the wave equation for fear of straying too far from experimental evidence? This is an interesting speculation. Moreover, in the rest of his chapter, de Broglie discusses the classical  Rayleigh scattering theory of light and writes:
\begin{quote}
Qualitatively, the phenomenon will be analogous to the scattering of $\alpha$ and $\beta$ particles. The homogeneous wave that statistically represents the movement of [light] quantas propagates as if there were an index of refraction around the scattering center. But this homogeneous wave is nothing other than the wave of classical theories and, from the nature of the validity attributed by us to these theories, we must believe that they will still give us here an exact global representation of this diffusion. \cite{debroglie1925b}
\end{quote}
All this shows that for de Broglie, the analogy between light and matter is perfect. He therefore had no reason to doubt the strength of his general theory based on a $\Psi$ wave of local index $n_E(\mathbf{x})$, even outside the semi-classical limit. All of which goes to show that de Broglie had a very accurate physical understanding of the underlying physics, and of course it makes it all the stranger that the wave equation is not formally explained.
We find ourselves in the curious situation of a historian desperately looking for Newton's formula $F=m\gamma$ in the famous \textit{Principia}, even though it was later introduced by Euler! 
Perhaps that's where the explanation lies. Schr\"odinger was a more experienced physicist than Louis de Broglie, and as many historians have pointed out, he had no trouble filling de Broglie's mathematical gaps.  In confirmation of this point of view, I would like to mention the interview Louis de Broglie gave to Fritz Kubli in 1968 (reproduced in \cite{Kubli} and discussed in \cite{Kubli2}). I will reproduce here a short extract from the interview:
\begin{quote}
F. K.: What interests me is what you said in another interview, that you didn't know at the time [1926] that Fermat's principle is a consequence of the wave equations by passage to the limit!

L. de Broglie: Yes - perhaps I didn't know the demonstration. I knew that one could deduce the equation of geometrical optics from the wave equation, but I don't know if I had the demonstration in mind - which I later gave in a number of my books. Probably I didn't know it, or didn't know it until Schr\"odinger's time or thereabouts.

F. K.: And I think this demonstration is very important because Schr\"odinger took it up again.

L. de Broglie: Yes, that's it, that's it [...]. \cite{Kubli}
\end{quote}

The interview goes on to discuss Courant and Hilbert's mathematical work, which is a reference for theoretical physicists and discusses wave theory in particular. The fact that de Broglie was unaware of this work in 1925 was for him a definite mathematical lacuna. Similarly, he was unaware of the details of the Hamilton-Jacobi theory used by Schr\"odinger. He made up for this after the event, making extensive use of Hamilton-Jacobi theory in conjunction with Madelung's hydrodynamic theory, also published in 1926, to form the basis of the double solution and pilot wave theories. It's also important to note that our analysis concerns only the case of stationary theory (i.e. at constant $E$ energy), for a single particle. The many-body case ($N$ electrons) was also correctly guessed by Schr\"odinger within the framework of the $3N$-dimensional configuration space, which lends itself well to the generalization of the classical Hamilton-Jacobi equation. The configuration space is also a key element of the pilot--wave theory.

At the very end of his book Ondes et mouvements, de Broglie (during the proofreading stage)  added that he just read the  Schr\"odinger's articles  \cite{Schrodinger1926} (that the author actually sent to him). De Broglie then summarized the content of Schr\"odinger work by introducing precisely the derivation that Darrigol mentioned in his article \cite{Darrigol2013} (a derivation that is, incidentally, simpler than Schr\"odinger's in his original articles). This completes the circle, but is only the beginning of the quantum story, as we all know.

\vskip 30pt
\begin{eref}
\bibitem{debroglie1923a}
 L. de Broglie, Ondes et quanta.  C.~R. Acad.~Sci. (Paris), \textbf{177}, 507--510 (1923).
\bibitem{debroglie1923b}
L. de Broglie, Quanta de lumi\`{e}re, diffraction et interf\'erences.  C.~R. Acad.~Sci. (Paris), \textbf{177}, 548--560 (1923). 
\bibitem{debroglie1923c}
L. de Broglie, Les quanta, la th\'eorie cin\'etique des gaz et le principe de Fermat.  C.~R. Acad.~Sci. (Paris), \textbf{177}, 630--632 (1923).
\bibitem{debroglie1925th}
L. de Broglie, Recherches sur la th\'eorie des quanta, Facult\'e des Sciences de Paris (1924); Annales de Physique  l0-\`eme s\'erie, \textbf{III}, 22--128 (1925).
\bibitem{Schrodinger1926}
E. Schr\"odinger, Quantisierung als eigenwertproblem, Ann. der Phys. \textbf{384}, 361-376; \textbf{384}, 489-527; \textbf{385}, 437-490; \textbf{386}, 109-139  (1926).
\bibitem{Forman}
V. V. Raman and P. Forman, Why was it Schr\"odinger who developed de Broglie's ideas?, Historical Studies in the Physical Sciences, \textbf{1}, 291-314 (1969).
\bibitem{Darrigol2013}
O. Darrigol, A few reasons why Louis de Broglie discovered matter waves and yet did not discover Schr\"odinger's equation, in Erwin Schr\"odinger 50 years after Edited by W. Reiter et al., European Mathematical Society, 166-174 (2013).
\bibitem{debroglie1925}
L. de Broglie,  Sur la fr\'equence propre de l'\'electron, C.~R. Acad.~Sci. (Paris), \textbf{180}, 498--500 (1925).
\bibitem{debroglie1926}
L. de Broglie, Les principes de la nouvelle m\'ecanique ondulatoire, J. Phys. Radium \textbf{7}, 321-337 (1926).
\bibitem{debroglie1927}
L. de Broglie, La m\'ecanique ondulatoire et la structure atomique de la mati\`ere et du rayonnement, J. Phys. Radium \textbf{8}, 225-241 (1927). 
\bibitem{debroglie1956}
L. de Broglie, Une tentative d'interpr\'etation causale et non lin\'eaire de la m\'ecanique ondulatoire: la th\'eorie de la double solution, Gauthier-Villars, Paris, France, (1956);  translated  as Nonlinear wave mechanics: A causal interpretation, Elsevier, Amsterdam, Netherlands  (1960).
\bibitem{Fargue} 
D. Fargue, Permanence of the corpuscular appearance and non linearity of the wave equation, in The wave-particle dualism edited by S. Diner et al. D. Reidel Publishing, Dordrecht, Netherlands, 149-172 (1984).
\bibitem{Drezet2021} 
A. Drezet,  The guidance theorem of de Broglie. Ann. Fond. de Broglie, \textbf{46}, 65--85 (2021).
\bibitem{Drezet2023}
A. Drezet,  Quantum solitodynamics: Non--linear wave mechanics and pilot--wave theory, Found. Phys. \textbf{53}, 31 (2023).
\bibitem{Drezet2023b}
A. Drezet,  A time-symmetric soliton dynamics \`{a} la de Broglie. Found. Phys. \textbf{53}, 72 (2023).
\bibitem{Drezet2024}
A. Drezet, Whence nonlocality? Removing spooky action-at-a-distance from the de Broglie Bohm pilot--wave theory using a time-symmetric version of the de Broglie double solution, Symmetry \textbf{16}, 8 (2024).
\bibitem{Valentini}
G. Bacciagaluppi, A. Valentini, Quantum theory at the crossroads: Reconsidering the 1927 Solvay Conference, Cambridge University Press, Cambridge, UK (2009). 
\bibitem{debroglie1930}
L. de Broglie, Introduction \`{a} l'\'{e}tude de la m\'{e}canique ondulatoire, Hermann, Paris, France (1930). English translation as An introduction to the study of wave mechanics, Methuen, London, UK (1956). 
\bibitem{Bohm1952}
D. Bohm, A suggested interpretation of the quantum theory in terms of hidden variables-I and II, Phys. Rev. \textbf{85}, 166-179; 180-193 (1952).
\bibitem{debroglie1926b}
L. de Broglie, Remarques sur la nouvelle m\'ecanique ondulatoire,  C.~R. Acad.~Sci. (Paris), \textbf{183}, 272--274 (1926).
\bibitem{debroglie1925b}
L. de Broglie, Ondes et mouvements, Gauthier-Villars, Paris, France (1926). 
\bibitem{Shanahan2024}
D. Shanahan, The de Broglie wave as an undulatory distortion induced in the moving particle by the failure of simultaneity, to appear  in Ann. Fond. de Broglie, \textbf{48}, (2024).
\bibitem{Kubli}
F. Kubli, Un entretien avec Louis de Broglie (20 Novembre 1968), Ann. Fond. de Broglie, \textbf{17}, 111--134 (1992).
\bibitem{Kubli2}
F. Kubli, Conversation avec Louis de Broglie au sujet de sa th\`ese, in La d\'ecouverte des ondes de mati\`eres, Acad\'emie des Sciences, 55--64 (1994).

\end{eref}
\end{document}